# Induced surface enhancement in coral Pt island films attached to nanostructured Ag electrodes

H. Khoa Ly[1], Christopher Köhler[2], Anna Fischer[1], Julia Kabuß[2], Felix Schlosser[2], Mario Schoth[2], Andreas Knorr[2], Inez M. Weidinger[1]*

[1]Technische Universität Berlin, Institut für Chemie, Straße des 17. Juni 135, D-10623 Berlin, i.weidinger@mailbox.tu-berlin.de, Germany. [2]Technische Universität Berlin, Institut für Theoretische Physik, Hardenbergstraße 36, D-10623 Berlin, Germany

Coral Pt islands films are deposited via electrochemical reduction on silica coated nanostructured Ag electrodes. From these devices surface enhanced (resonance) Raman [SE(R)R] signals of molecules exclusively attached to Pt are obtained with intensities up to 50% of the value determined for Ag. SE(R)R spectroscopic investigations are carried out with different probe molecules, silica coating thicknesses and excitation lines. Additionally, field enhancement calculations on Ag-SiO$_2$-Pt support geometries are performed to elucidate the influence of the Pt island film nanostructure on the observed Raman intensities. It is concluded that the non perfect coating of the Pt island film promotes the efficiency of the induced Pt SER activity. Comparison with similar measurements on Ag-SiO$_2$-Au electrodes further suggests that the chemical nature of the deposited metal island film plays a minor role for the SE(R)R intensity.

KEYWORDS Surface chemistry, surface enhanced Raman spectroscopy, Platin, Plasmons, metal island films

## 1. Introduction

Surface chemistry plays a key role in many fields of modern technology such as heterogeneous catalysis, solar energy conversion, or bioelectronics. The efficiency of the surface-confined processes sensitively depends on the interaction of adsorbates with their underlying support. Hence, changing the chemical nature and surface morphology of the support as well as the properties of the surrounding medium can greatly enhance or diminish the performance of the device. Therefore considerable research efforts have been made to design new types of support materials. However, probing the adsorbates and their processes on such surfaces yet represents a considerable challenge since there are no generally applicable analytical techniques that provide information about the molecular structure and dynamics of the adsorbates under *in situ* conditions.

In contrast to most surface-sensitive methods that are only applicable to solid/gas interfaces at very low pressures, surface enhanced Raman spectroscopy (SERS) can be employed to surfaces irrespective of the kind of the surrounding medium and thus may be the in-situ analytical method of choice in surface chemistry.[1-5] The main drawback of SERS that currently prevents a wider applicability is related to the signal enhancement mechanism which requires the resonant coupling of radiation with surface plasmons of the metallic support. Such plasmon resonances strongly depend on the dielectric function and surface morphology of the metal, and so far mainly Ag and Au have been demonstrated to be capable of providing sufficient surface enhancement for reliable SER analysis.[6] Especially Ag affords a strong surface enhancement in a wide spectral range from the violet to the near-infrared region,[7-9] but unfortunately this metal is only of minor interest for technological applications. In this respect, other metals such as Pt or Pd are much more relevant, however, their intrinsic SERS activity is lower by several orders of magnitude than in the case of Ag[10] such that they are not considered as suitable supports for in situ SER spectroscopy.

To overcome this drawback several approaches have been developed in the past to induce SER activity for molecules adsorbed on plasmonic inactive surfaces. All these approaches have in common that they are based on hybrid systems consisting of a plasmonic component such as Ag or Au for optical amplification and a non-plasmonic component for surface chemistry: In tip enhanced Raman spectroscopy (TERS) surface enhancement is provided by a nanoscaled Au(Ag) tip that is brought in close vicinity to the adsorbate on a non-plasmonic support.[11-13] Although much insight can be given by this technique it requires a rather demanding setup. In-situ spectroscopy, however, does not necessarily need the high spatial resolution of TER spectroscopy but relies upon a flexible set-up that can easily adapted to the specific applications such as heterogeneous catalysis in reactors or electrochemical cells in bioelectronics. Therefore alternatively a strategy was proposed where the tip is replaced by an ensemble of nanoparticles that are spread onto the probed surface.[14-16] This approach avoids the sophisticated TERS setup but at the same time largely decreases the available area for surface chemistry.

Finally an overlayer approach was introduced by Weaver et al. and further developed by us and others where surface enhancement is induced into a non-plasmonic metal film by an underlying nanostructured SERS active support.[17-23] In our previous work, electrochemically roughened Ag electrodes were used as SERS active support and Au as outer metal layer for surface chemistry. The induced surface enhancement experienced by molecules attached to Au was measured under violet light excitation where Au itself does not show intrinsic plasmonic activity. In contrast to the strategy of Weaver et al. we introduced a defined separation of Au from the rough Ag electrode by coating the latter with a dielectric spacer of defined thickness. Repellent functionalisation of the spacer layer guaranteed exclusive adsorption of the adsorbates under investigation on Au even for a non perfectly closed metal film. With this strategy the enhancement of Raman signals of the probe molecule, i.e. the heme protein cytochrome c (Cyt-c), was found to be similar on the outer Au layer as for direct immobilisation on the Ag support. This highly unusual long range enhancement over a spacer thickness of more than 20 nm was rationalised by field enhancement calculations of Ag-spacer-Au devices.[24]

In this paper we have successfully expanded our approach to coral Pt island films. By variation of the excitation line, the target molecules and the spacer thickness, it was possible to identify crucial parameters for optimum induced surface enhancement at plasmonic inactive metal films. In addition, field enhancement calculations were performed on different multilayer Ag-spacer-Pt geometries to rationalize the experimental results.

## 2. Materials and Methods

### 2.1 Chemicals

Tetrachloroplatinate(II) acid ($H_2[PtCl_4]\cdot 6H_2O$, 99.9 %), tetraethyl orthosilicate (TEOS, 99.99 %), aminopropyl triethoxysilane (APTES), 3-Mercaptopropylmethyldimethoxysilane (MPTS), 11-amino-1-undecanethiol (AUT), and 11-mercapto-undecanoic acid (MUA) were purchased from Sigma Aldrich. Ethanol (99.99 %), isopropyl alcohol (99%) and ammonium hydroxide (35% aqueous solution) were obtained form Fischer Scientific Company (Germany). Potassium hydrogen phosphate and potassium dihydrogen phosphate were provided by Merck (Germany). Benzolthiol (BT, 98%) and mercaptopyridine (mPy, > 95 %) were purchased from Sigma-Aldrich. All reagents were of analytical grade and used as received. Solutions were prepared using ethanol of analytical grade (99.99%) or Millipore water (Eschborn, Germany) with a resistance > 18 MΩ. Ag ring electrodes of 8 mm diameter and 2.5 mm height were machined from 99.99% Ag rods (Goodfellow, U.K.). Cytochrome c (Cyt c, Sigma-Aldrich) was purified as described previously.[25]

*2.2 Spectroscopic and electrochemical measurements*

Spectro-electrochemistry was performed using cylindrical Ag and Ag-S-Pt rings as working electrode, an Ag/AgCl (3 M KCl) reference electrode (+0.21 V vs. SHE) and a platinum counter electrode. SER(R) spectra were measured using a confocal Raman spectrometer (LabRam HR 800, Jobin Yvon) coupled to a liquid nitrogen cooled CCD detector. The spectral resolution was 1 cm$^{-1}$ with an increment per data point of 0.28 cm$^{-1}$ and 0.15 cm$^{-1}$ using 413 nm and 514 nm laser excitation line, respectively. For laser excitation the 413 nm laser line of a Krypton- (Coherent Innova 300c) or the 514 nm line of an Argon cw-laser (Coherent Innova 70c) was used. The laser power on the sample was 1.0 mW. The laser beam was focused onto the sample by a Nikon 20x objective with a working distance of 20.5 mm and a numeric aperture of 0.35. Accumulation times of the SERR spectra were between 1 and 12 s. The working electrode was constantly rotated to average over a high electrode surface and to avoid laser induced sample degradation. All SER(R)S measurements were done in 30 mM PBS buffer solution. Cyclic voltammetric experiments were performed with a CH instrument 660 C (Austin, USA).

*2.3 Electrode characterization*

The surface morphology and elemental composition of the electrodes was characterized with scanning electron microscopy (SEM) and energy dispersive X-ray spectroscopy (EDX) using a JEOL 7401F operated between 8 and 12 kV equipped with an EDX detector Quantax XFlash® Detektor 4010 from Bruker. Specific surface area measurements by using multi point BET were done with Krypton gas adsorption measurements at 77,4 K with an Autosorb-1-C from Quantochrome. The samples were degassed at 80°C over night prior measurement.

*2.4 Theoretical calculations*

The field distribution was simulated using the Maxwellsolver JCMsuite, a finite element software for the computation of electromagnetic waves, developed by the Zuse Institut Berlin.[26] The experimental structures were modelled by a three-dimensional but rotationally symmetrical geometry. An external electromagnetic field was applied with defined polarization wavelength and vector amplitude to calculate the field enhancement.

**3. Results**

*3.1 Electrode preparation*

Ag-SiO$_2$-Pt electrodes were created following the procedure established for Ag-SiO$_2$-Au in our previous work [19;21] with some modifications. The steps for electrode preparation are illustrated in the top half of Figure 1: First, a nanostructured Ag support was created from a cylindrical Ag bulk electrode via electrochemical roughening using an established protocol.[27;28] The rough Ag electrode was then coated with a silica layer of variable thickness. Thin layers (0.7 nm) were formed by self assembly of MPTS in aqueous solution.[21;29] Thicker SiO$_2$ layers were generated by additional incubation of the electrode in a TEOS solution for 2 h.[30] The thickness of the SiO$_2$ layer could thereby be tuned by adjusting the concentration of the TEOS precursor.[31] Subsequently the Ag-SiO$_2$ electrode was dipped in APTES solution over night to obtain a positively charged amino-functionalised surface. This step resulted in an additional layer thickness of 1.0 nm.[32;33]

An SEM picture of an Ag-SiO$_2$ electrode surface is shown in Figure 1A. It can be clearly seen that a closed and dense SiO$_2$ film covers completely the Ag surface. The thickness of the SiO$_2$ film in Figure 1A was determined to be 15 nm corresponding to an initial TEOS concentration of 6 μM. It should be noted that attaching SiO$_2$ does not affect the surface morphology of the electrode. If the surface pattern of the Ag electrode is approximated by an arrangement of connected coral spheres an average coral diameter of d= 85 ± 20 nm can be estimated. In the last step, the Ag-SiO$_2$ electrode was dipped into an ethanol solution of 2% wt of H$_2$[PtCl$_4$]·6H$_2$O for 1 - 30 min to allow for electrostatic binding of PtCl$_4^-$ ions to the positively charged amino groups of the silica surface. After changing the buffer solution, a potential of -0.5 V was applied which resulted in a reduction of [PtCl$_4$]$^{2-}$ ions and a Pt island film was generated on top of the Ag-SiO$_2$ electrode.

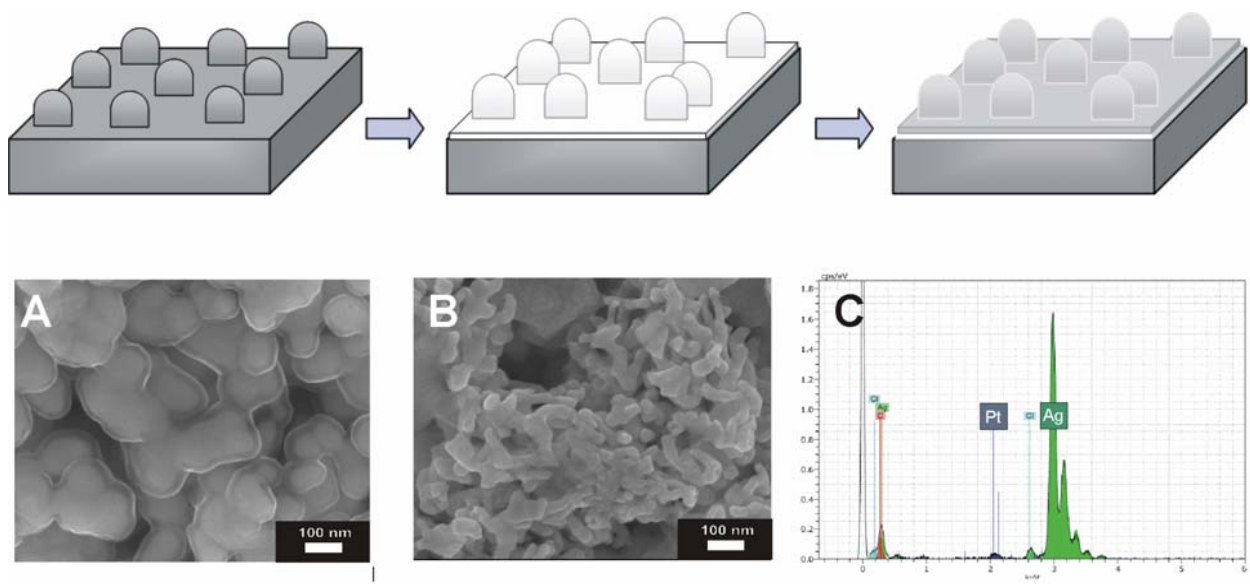

**Figure 1.** Schematic presentation of the Ag-SiO$_2$-Pt electrode preparation. First, an Ag electrode was electrochemically roughened, followed by coating with a dielectric SiO$_2$ spacer layer of defined thickness. Finally, a nanostructured Pt island film was electrochemically deposited on top of the spacer. Bottom: SEM pictures of Ag-SiO$_2$ (A) and Ag-SiO$_2$-Pt (B) electrodes. Corresponding EDX spectra of the Ag-SiO$_2$-Pt electrode (C).

In Figure 1B an SEM picture of the electrode after electrochemical Pt deposition is shown. The electrode surface is now covered by a Pt island film that exhibits also a coral like structure but with a smaller coral size than the underlying Ag support. The average diameter of the Pt nanocorals is approximated by 30 ± 10 nm. The presence of Pt could be proven by EDX measurements that can be seen in Figure 1C. It has to be noted that with increasing incubation time of the Ag-SiO$_2$ coated electrode in the PtCl$_4$ solution also islands with a thicker, film-like morphology were formed upon Pt reduction (see Figure S1 supporting information). Thicker and thus more closed films, however, led to a decrease in SER intensity of adorbates (*vide infra*).

*3.2 Determination of REF for rough Ag electrodes*

Prior to estimation of the surface enhancement of Ag-SiO$_2$-Pt electrodes the Raman enhancement factor (REF) of the pure electrochemically roughened Ag electrodes had to be determined according to Eq. 1

$$REF = \frac{I_{SERS}}{I_R} \cdot \frac{N_R}{N_{SERS}} = \frac{I_{SERS}}{I_R} \cdot \frac{c_R \cdot V}{\Gamma_{SERS} \cdot A} \qquad (1)$$

where $I_R$ and $I_{SERS}$ are the Raman intensities of probe molecules in solution and adsorbed on the Ag surface, respectively. These intensities are normalised to the same accumulation time and laser intensity. $N_R$ and $N_{SERS}$ refer to the number of molecules that are in the focus of the laser beam. These quantities are related to the product of illuminated volume V and the bulk concentration $c_R$ in the normal Raman- and to the product of the illuminated area A and the surface concentration $\Gamma_{SERS}$ in the SER experiments. Benzothiol (BT) was used as a Raman probe to estimate REF following the procedure described by McFarland et al.[34] Raman and SER spectra were measured for BT in solution and adsorbed on the surface using 413 nm and 514 nm excitation. The intensity ratio of $I_{SERS}/I_R$ was subsequently determined for the 1000 cm$^{-1}$ and 1080 cm$^{-1}$ line. The average ratio considering both bands is shown in Table 1. The concentration of the neat BT solution was $c_R$ = 9.54 mol/L. The irradiated volume V was approximated by a cylinder with radius r and height h

$$V = \pi \cdot r^2 \cdot h \qquad (2)$$

r is given by the radius of the laser focus which has been estimated by McFarland et al. to be r = 2 μm for a 20x objective independent of the laser wavelength. The height of the cylinder can be approximated by the depth of focus of a Gaussian beam:

$$h = 2 \cdot r^2 \cdot \frac{\pi}{\lambda} \qquad (3)$$

Accordingly the surface area probed in the SERS measurements is given by:

$$A = RF \cdot A_{geom} = RF \cdot \pi \cdot r^2 \qquad (4)$$

$A_{geom}$ stands for the geometrical area of the electrode illuminated by the laser. To obtain the real surface area one has to multiply by a factor that accounts for the surface roughness of the electrode. The surface roughness factor (RF) was determined independently by BET measurements to be 20. The surface concentration of $\Gamma_{SERS}$ = 1.1 nM/cm$^2$ of BT on Ag was taken from ref. [34]

On the basis of these parameters, we obtain a value for REF = 2·10$^3$ for 514 nm and 5·10$^2$ for 413 nm excitation. These values have to be seen as a lower limit as we assume that all parts of the surface are accessible by laser light and that a compact BT monolayer is formed on the surface. The nevertheless low REF values can be rationalized by the fact that, due to the random coral structure, a large fraction of the surface plasmons is not in resonance with the incoming light. Hence the averaged signal that is observed in the SERR spectrum might originate only from a small percentage of the surface area. In return this heterogeneity ensures surface enhancement over a broad range of excitation lines.

**Table 1:** Ratio between SERS and normal Raman intensities of BT and corresponding Raman enhancement factors (REF) of electrochemically roughened Ag electrodes.

| λ / nm | $I_{SERS}/I_R$ | REF |
|---|---|---|
| 514 | 1.1 | $2 \cdot 10^3$ |
| 413 | 0.2 | $5 \cdot 10^2$ |

*3.3 SER spectroscopy of mPy on Pt*

Mercapto-pyridine (mPy) was used as a Raman probe to test surface enhancement at the Pt surface in Ag-SiO$_2$-Pt hybrid electrodes. SER spectra of mPy adsorbed on Ag (black) and Pt (grey) at 514 nm excitation are shown in Figure 2 A. The top spectrum refers to a SiO$_2$ spacer thickness of 1.7 nm, corresponding to a MPTS-APTS layer. The overall intensity of this spectrum is lower by a factor of 2 compared to the SER spectrum of mPy adsorbed on Ag (bottom). The top spectrum in Figure 2A can be exclusively attributed to mPy molecules that are adsorbed to the Pt surface as mPy is expected to bind via its thiol group covalently to metal surfaces only. Contributions from mPy directly attached to the Ag support can be ruled out since no SER signal was detected for the Ag-MPTES-APTES electrode prior to Pt deposition. The intensity of the 1096 cm$^{-1}$ band of mPy adsorbed on Ag was ascribed to the Raman enhancement factor (REF) of $2 \cdot 10^3$ (*vide supra*). Correspondingly, REF for mPy on Pt was derived by comparison of this band's relative intensity in respect to its value on Ag. The values for log(REF) obtained in this way are plotted in Figure 2 B for two different spacer thicknesses, assuming the same mPy surface coverage. The data show that REF decreases with spacer thickness, supporting the view that the surface enhancement generated at the Ag surface is mainly responsible for the observed SER intensity. Nevertheless, even for a 10 nm thick spacer, REF has decreased only by a factor of ca. 5 compared to bare Ag. The decrease in REF as a function of SiO$_2$ thickness may be considered to be linear with a slope of $\Delta \log(REF)/\Delta d = 0.06$ nm$^{-1}$.

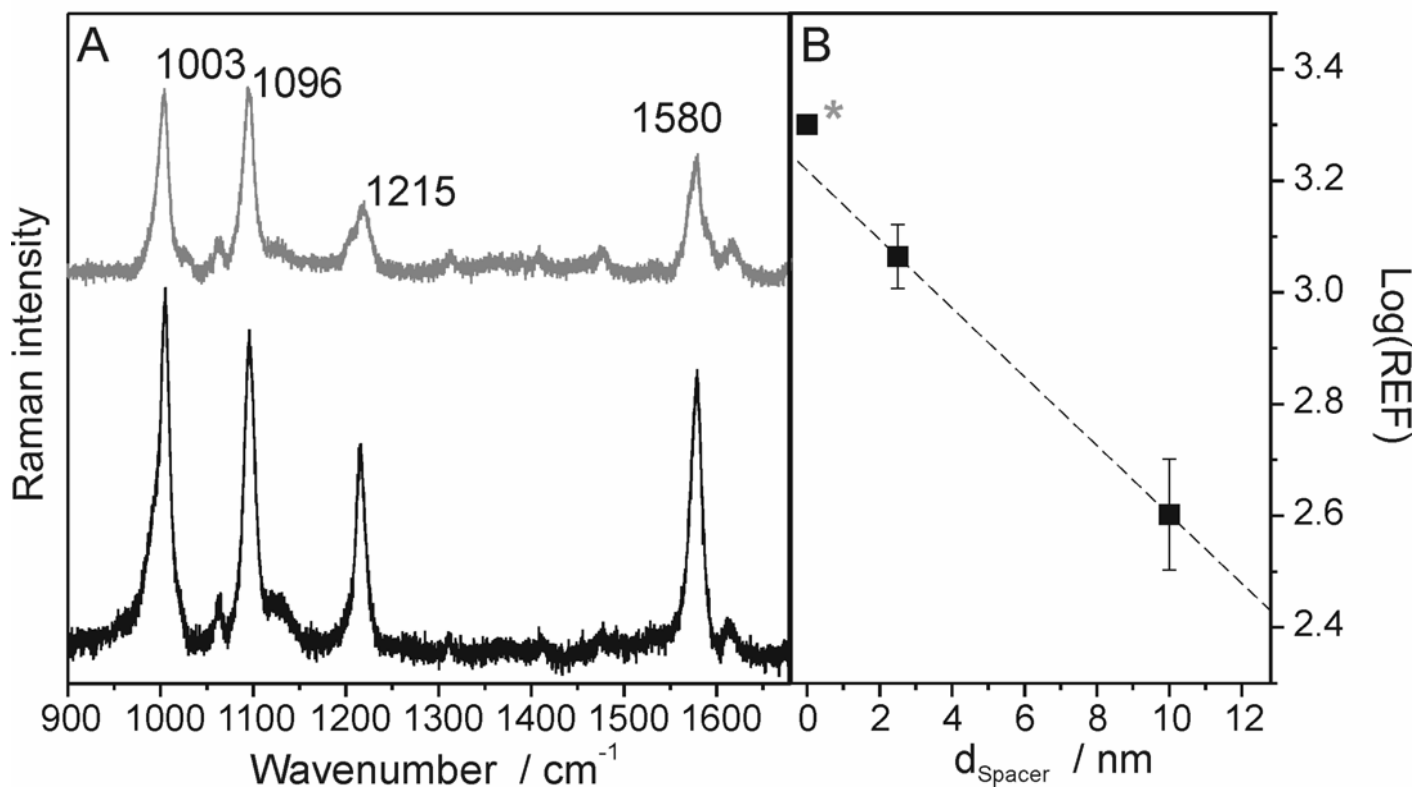

**Figure 2.** (A) SER spectra of mPy attached to Ag (black) and Ag-MPTS-APTES-Pt (grey) surfaces. (B) Raman enhancement factors (REF) of mPy attached to Pt as a function of the spacer thickness. The value marked with an asterix corresponds to mPy on bare Ag surfaces.

*3.4 SERR spectroscopy of Cyt c at Pt*

Intense SERR (surface enhanced resonance Raman) spectra are obtained for the heme protein cytochrome c (Cyt c) attached to SER active electrodes under violet light excitation which matches the electronic transition of the heme chromophore. Furthermore, the redox properties of Cyt c are preserved if the metal electrode is coated by biocompatible self assembled monolayers (SAM). Cyt c is therefore widely used as a reference system to test surface enhancement and electrical communication of novel nanostructured electrodes.[21;35]

Cyt c directly adsorbed on bare Ag surfaces is known to undergo a partial transition to a non-native high spin state of the heme which can be distinguished from its native state in the SERR spectrum.[36;37] To avoid this adsorption-induced protein denaturation the electrodes are usually coated with a SAM of ω-carboxyl alkanethiols. The increased biocompatibility, however, is achieved at the expense of surface enhancement that drops i.e. by a factor of 2 after coating with SAMs of mercaptoundecanoic acid (MUA). For Cyt c directly adsorbed to the Pt surface in Ag-SiO$_2$-Pt electrodes SERR spectra with high intensity could be obtained.

However, the spectra also indicated a strong contribution of the non-native high spin state (Figure S2 Supporting information), accompanied by a time dependent decrease in SERR intensity. In order to obtain stable SERR signals, the Pt surface had to be coated with MUA prior to Cyt c adsorption, in analogy to previous experiments on Ag surfaces. The spectra obtained this way were stable over time and could be quantitatively described by a superposition of the spectra only of the native ferrous and ferric form of Cyt c. The SERR spectra of Cyt c attached to Ag-MUA and Ag-S-Pt-MUA with S = MPTES-APTES differ only slightly in SERR intensity which drops by roughly a factor of 2 for the Pt hybrid system (Figure 3A). By changing the applied potential of the spectro-electrochemical cell the protein could be completely reduced and re-oxidized when adsorbed on the Pt surface which demonstrates that the electrical communication between the working electrode and Cyt c is still intact. The Cyt c midpoint potential was determined to be $E_m$ = 20 mV which is identical to its value on Ag-MUA.[38]

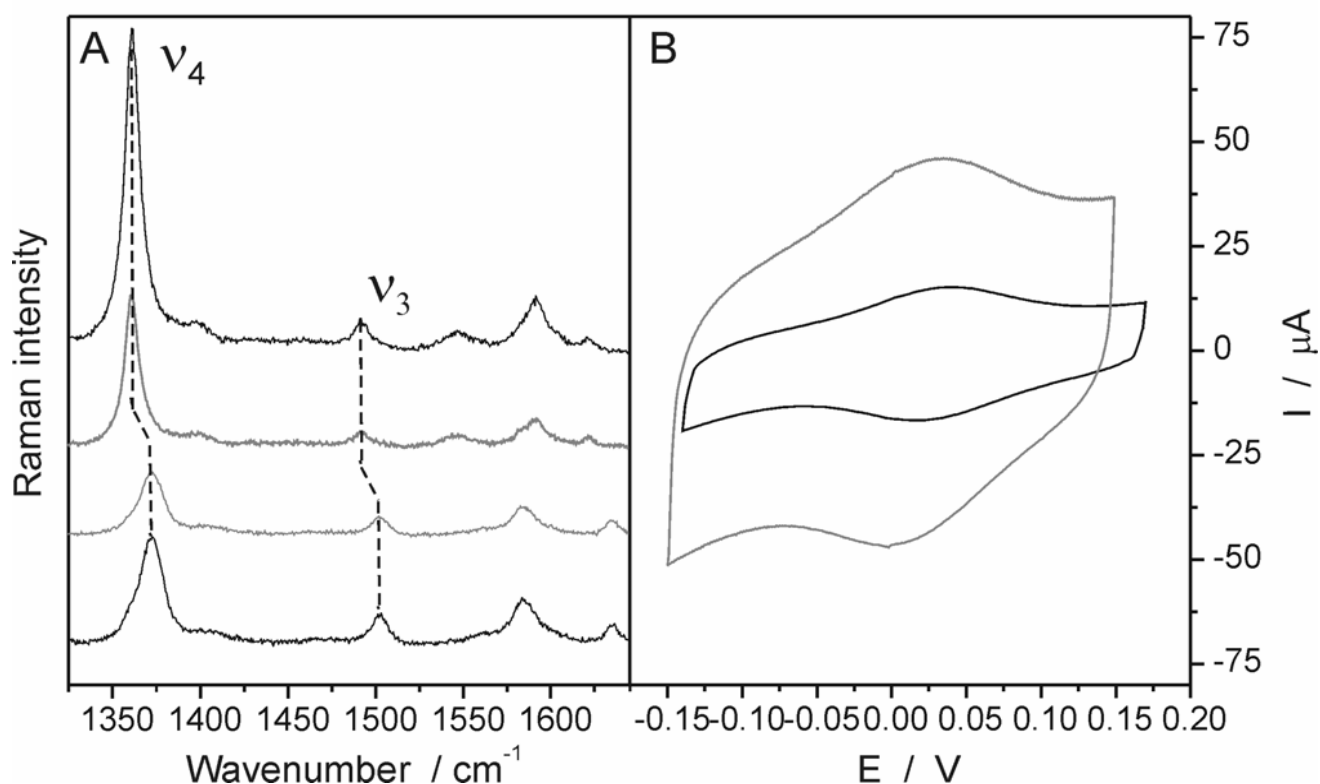

**Figure 3.** (A): Potential dependent SERR spectra of Cyt c on (from top to bottom): Ag-MUA (E=+150 mV), Ag-S-Pt-MUA (E=+150 mV), Ag-S-Pt-MUA (E=-400 mV), Ag-MUA (E=-400 mV). (B): CV of Cyt c on Ag-MUA (black) and Ag-S-Pt-MUA (grey). S= MPTES-APTES, Scan rate: 100 mV/s.

REF on the Pt surface was obtained by comparison of the relative Cyt c SERR intensities attached to Ag-MUA and Ag-S-Pt-MUA supports. Thus, the REF on Ag-MUA was estimated to be $2.5 \cdot 10^2$ taking into account the effect of the MUA coating on surface enhancement (*vide supra*). In order to eliminate the effect on signal intensity that arises from variations in the number of probed Cyt c molecules CV measurements were carried out concomitant to the spectroscopic investigations. CV plots of Cyt c attached to either Ag-MUA or Ag-S-Pt-MUA electrodes are shown in Figure 3 B. Integration of the voltammetric peaks allowed to determine the Cyt c surface coverage $\Gamma_{Cyt}$ for both systems. $\Gamma_{Cyt}$ was calculated for Ag-MUA to be $6.5 \cdot 10^{-11}$ M cm$^{-2}$ in respect to the geometrical area of the electrode. On Ag-S-Pt-MUA a nearly 2 times higher value of $1.2 \cdot 10^{-10}$ M cm$^{-2}$ was determined. One has to note that these values represent apparent surface concentrations as the respective surface roughness of the different electrodes is not considered. Multiplication of $\Gamma_{Cyt}$ with the area of the laser spot gives the number of Cyt c molecules that contribute to the observed Raman signal. Thus REF on Pt was corrected for the higher number of probed Cyt c molecules and is presented in Figure 4 A as a function of SiO$_2$ spacer thickness. In Figure 4 A also the REF values previously determined for Ag-SiO$_2$-Au-MUA electrodes[21;39] are shown. As can be seen the overall value for REF is slightly higher if one uses Au as outer metal. Nevertheless the slope of a linear fit in Figure 4 A, which was determined to be $\Delta \log(REF)/\Delta d = 0.025$ nm$^{-1}$ for Pt coatings, is nearly the same for Au and Pt hybrid electrodes.[39]

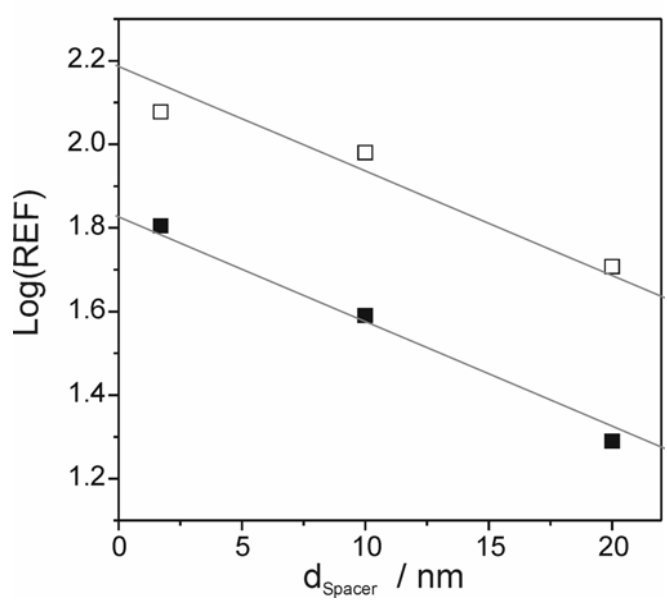

**Figure 4.** Raman enhancement factors (REF) for Cyt c on Ag-SiO$_2$-Au-MUA (open squares, taken from ref.[32]) and Ag-SiO$_2$-Pt-MUA (solid squares, this work) electrodes as a function of SiO$_2$ spacer thickness.

*3.5 Calculations*

The electric field enhancement distribution was calculated using finite element method for a electrode model geometry. The multilayered electrode depicted in Figure 5 A was modelled as follows: Starting with an infinite long bulk Ag electrode (y-direction) with a height of 100 nm, 3 half spheres were attached with a radius $r_0$ = 42.5 nm corresponding to the experimental determined average Ag coral size. The distance between the half spheres is unevenly distributed with two spheres being in close vicinity (6 nm) and a third one in a wider distance (62.5 nm). This structure is coated with a 2 nm dielectric spacer of $SiO_2$ and a 5 nm thin Pt film of same surface morphology. Water was taken as the surrounding medium. The applied external electromagnetic field is taken as a plane wave incoming from the right, propagating along the x direction. The light is assumed to be linearly polarized parallel to the y axis, the wavelength is set to λ = 413 nm and the vector amplitude is set to $|\vec{E}_0| = 1$. For the chosen geometry it can be clearly seen that surface enhancement of the fields is still present at the Pt-$H_2O$ interface. Due to the choice $|\vec{E}_0| = 1$ the field enhancement is given by the absolute value of the electric field $|\vec{E}|$. Maximum field enhancement is achieved in the interspaces of the two close-by half spheres. The perfect coated uniform Pt film in Figure 5A (2D plot left side), however, does not resemble the Pt film that is present in the experiments. One of the main differences compared to the real samples lies in the fact that the Pt island film, as shown in Figure 1, is not completely covering the underlying electrode, i.e. it exhibits hole-like defects. To approximate the experimental conditions more closely, holes were introduced into the Pt film shown in Figure 5 B. Interestingly; these defects enhance the average electric field at the Pt surface. Figure 5 (right side) shows the corresponding 3D plots of the field enhancement for the perfect and disordered system. It can be seen that introducing hole-like defects generates more "hot spots" for surface enhancement in the $SiO_2$/Pt interface. Most remarkable are the "hot spots" at the top of the half sphere and in the wider gap between the two spheres where no surface enhancement was observed in the absence of defects. It has to be further noted that a particularly large enhancement is observed at the sharp edges of the Pt islands, which are missing in closed films.

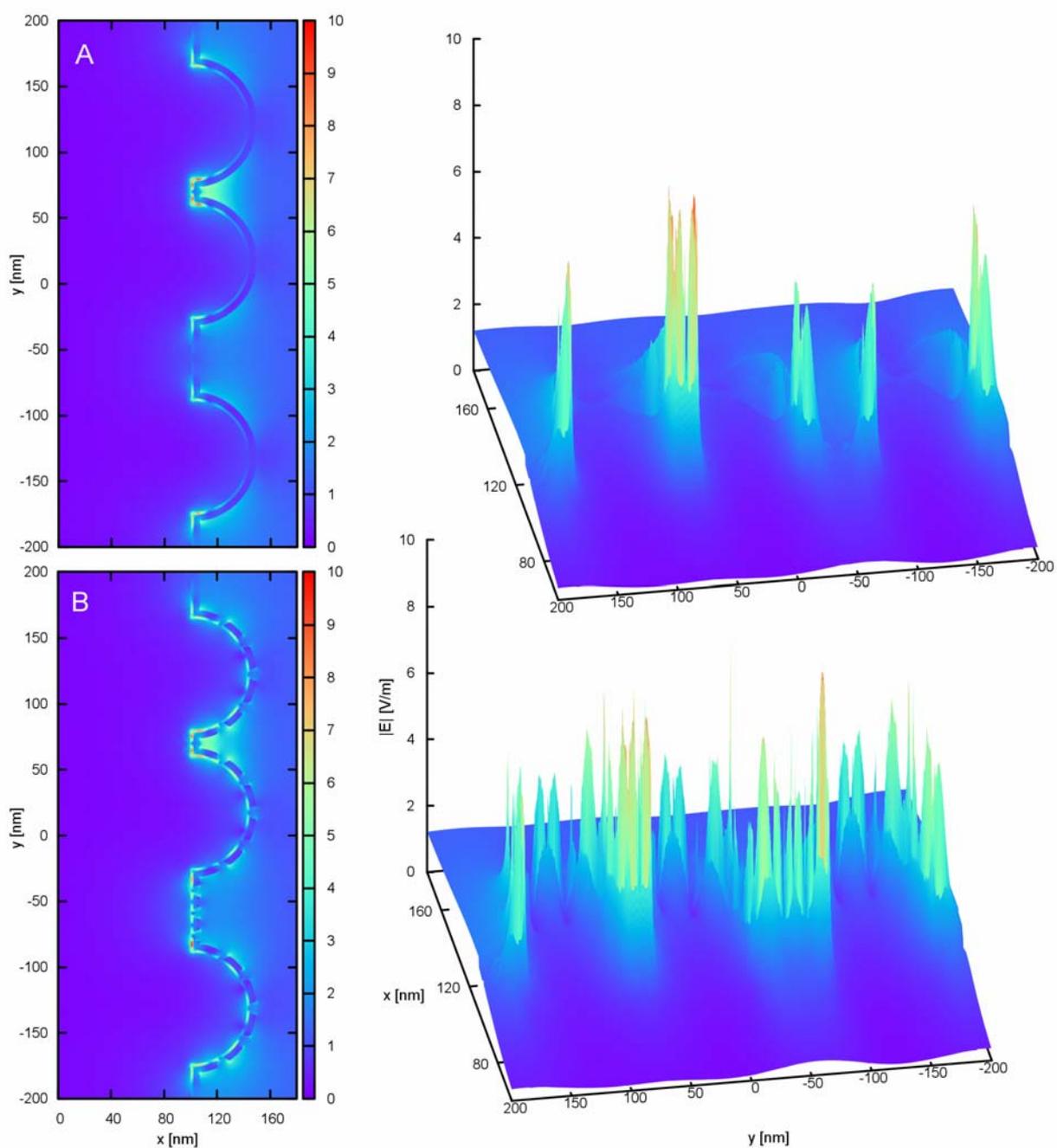

**Figure 5.** Field enhancement calculations of Ag-SiO$_2$-Pt geometries using a defect free (A) and defect containing (B) Pt film. On the right side the corresponding 3D plots of the field enhancement are displayed.

To obtain the average SERS enhancement for molecules adsorbed on Pt, $|\vec{E}|$ was read out at equidistant points 1 nm above the Pt surface for the geometries shown in Figure 5 A and B respectively. The SERS enhancement of adsorbates placed at these positions is then given by $g = |\vec{E}|^4$.[40] Direct comparison of both type of Pt films shows that the average SERS enhancement per surface area is increased by approximately 80% for the disordered film.

## 4. Discussion

High quality SER(R) signals in Ag-SiO$_2$-Pt constructs could be seen for non-resonant Raman (mPy) and RR probes (Cyt c) suggesting that the optical properties of the adsorbate itself do not play a role for the magnitude of induced surface enhancement at the Pt surface. Furthermore the comparable SERR enhancement, achieved for Pt and Au films in similar hybrid systems indicates that the chemical nature of the outer metal island film is also not highly relevant for the magnitude of induced SER activity. These results suggest that the *in situ* Raman detection method, proposed in this work, most likely can be applied for a variety of metal films and adsorbates.

The surface enhancement of rough Ag electrodes was determined to be 4 times higher at 514 nm- compared to 413 nm laser excitation. The same wavelength dependent enhancement ratio was observed on Pt if thin spacer layers (1.7 nm) were used. This finding supports our hypothesis that the optical properties solely of Ag are responsible for the observed enhancement at the outer metal film. However, the distance-dependent decrease of the REF was 3 times stronger for 514 nm excitation than for 413. It might be that different spots or local areas in the coral like structures of Ag and Pt are responsible for field enhancement at 514 and 413 nm. For thinner coatings the wavelength dependent field enhancement distribution is not altered whereas thicker coatings, that have a stronger influence on the overall surface morphology, shift the enhancement distribution to lower wavelengths.

This raises the question whether geometrical parameters of the multilayer hybrid system in general are the main factor for the magnitude of SER/SERR enhancement at the Pt surface.

The morphology of the Pt island film (Figure 1) shows a coral like nanostructure with smaller dimensions than the corals of the underlying Ag. Furthermore, the Ag-SiO$_2$ substrate contains areas with no or non detectable Pt. Both effects were simulated in the field enhancement calculations shown in Figure 5 by introducing small holes into the Pt film. As a result the average SERS enhancement is increased by ca. 80%.

This finding can be rationalised on the basis of two factors. First, the presence of holes will increase the number of incident photons on the Ag surface for excitation of surface plasmons and thus for field enhancement, whereas for closed Pt film losses due to reflection are much more severe. Second, the defects also create sharp edges within the Pt nanostructure.

According to theoretical predictions, such a geometric anisotropy in nanostructures provides a non-negligible contribution to the overall SER enhancement.[41;42] This so-called lightening rod effect focuses electric fields at the tip of metallic ellipsoids constituting an additional contribution to the field enhancement generated by metal particle plasmons. For Pt, which displays only a poor field enhancement due to particle plasmon resonances, such geometrical aspects are considered as the most important parameter for its marginal intrinsic SER activity.[10]

In view of these considerations, we conclude that the morphology of the Pt film plays the most crucial role for its observed high SER activity. Simple electrochemical deposition, as proposed in this work, creates Pt islands on top of coated Ag nanostructures with a favourable geometry for induced surface enhancement. This hypothesis is further supported by SERR measurements on Ag-$SiO_2$-Au hybrid electrodes where the outer Au film was formed via sputtering and thus exhibited a largely defect-free surface morphology. For this multilayer system no SERR signals of Cyt c could be detected.

## 5. Conclusion

This work presents a detailed investigation on how SER spectroscopy can be optimized in thin metal films that do not have intrinsic plasmonic activity. In the present Ag-spacer-Pt devices the SER intensity reaches ca. 50% of its value for Ag under otherwise identical conditions. The separation of Pt from Ag by a dielectric spacer has only a small effect on the surface enhancement but the silica coating efficiently prevents interaction of Ag with the adsorbates. As the most crucial parameter for the magnitude of SER enhancement the surface morphology of the metal island film could be identified whereas the chemical nature of the metal seems to play only a minor role. The defect containing outer Pt layer allows an efficient excitation of the Ag plasmons that are required to enhance the electric field in the vicinity of the Pt surface. The anisotropic shape of the Pt corals additionally promotes surface enhancement via the lightning rod effect. In summary, our results suggest that geometrical parameters of such multilayered electrode dominate the induced SERS efficiency on intrinsically SER-inactive metal films. Hence, the present approach should be applicable for analysis of adsorbates also on other types of metals, and thus may contribute to establish SER spectroscopy as a versatile *in situ* analytical tool for various applications.

ACKNOWLEDGMENT

The authors thank Peter Hildebrandt for encouragement and support. Furthermore we thank Ralph Krähnert and Ulrich Gernert for their support in SEM measurements and the Konrad-Zuse-Zentrum Berlin for providing the Maxwell solver JCMsuite (www.jcmwave.com). Financial support from the Fonds der Chemie and the DFG (Unicat, SPP Nanooptics) is gratefully acknowledged.

SUPPORTING INFORMATION

Further electrode characterization by SEM and SERRS. SERR spectra of Cyt c directly on Pt.

TOC

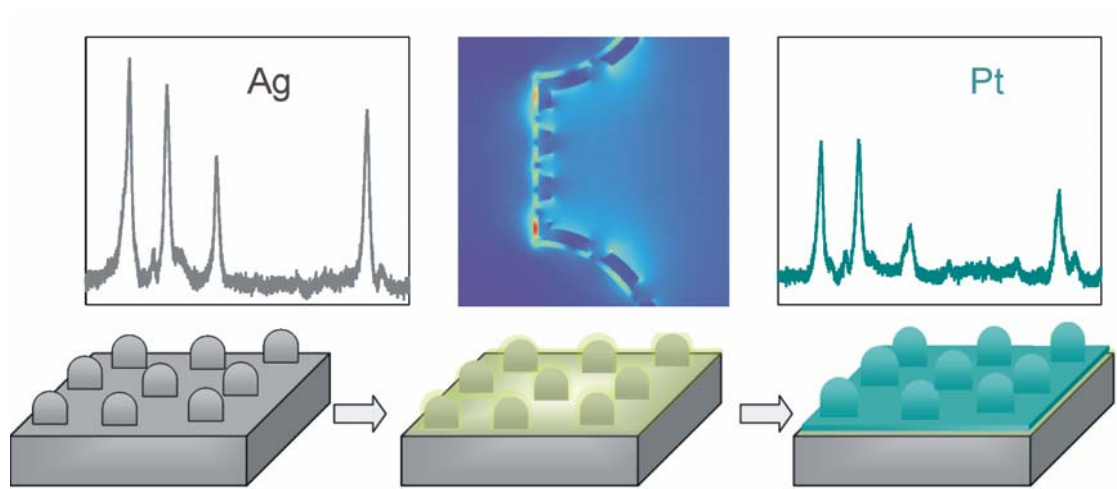